\documentstyle [11pt] {article}
\begin {document}

\title {\bf Motion of particles in storage rings in the presence of a
counterpropagating laser beam I}
\author {\it E.G.Bessonov \\
\small \it P.N. Lebedev Physical Institute RAS, 117924, Leninsky
prospect 53, \\
\small \it Moscow, Russia \\ bessonov@sgi.lpi.msk.su}
\date {} \maketitle

                       \begin {abstract}
The dynamics of charged particle beams interacting with laser beams and
material targets in storage rings is presented. Formulas for the
radiative damping, quantum excitation and equilibrium parameters of ion
and electron beams determined by the backward rayleigh and backward
compton scattering are derived. General case is considered. It takes
into account the possibility of the periodical displacement of the
laser beam in the radial direction with varying velocity in the process
of the interaction of the beams and includes the possibility of the
description of dynamics of muon beams interacting with material targets
in storage rings (enhanced ionization cooling). The example is
considered and some applications of being cooled beams are discussed.
\end {abstract}


\maketitle
                     \section {Introduction}

The backward compton or rayleigh scattering of laser photons by
elementary or complicated particles (electrons, protons, ions) in
storage rings results simultaneously both to radiative damping and to
quantum excitation of amplitudes of betatron and phase oscillations of
these particles. There is an analogy of these processes with the
influence of synchrotron radiation on dynamics of particles in storage
rings.

The spectral-angular characteristics of radiation scattered by an ion
and electron beams are identical. Therefore the equilibrium transverse
emittances and damping times of amplitudes of betatron and phase
oscillations of particle beams as a function of parameters of the
storage ring, charge and mass of particles are identical. However there
is a difference in the coefficients of proportionality of the
longitudinal emittances. In the relativistic case the maximal hardness
of scattered radiation according to the relativistic doppler effect is
$ 4 \gamma ^2 $ times higher then the hardness of the laser radiation,
where $ \gamma = 1/ \sqrt {1 - \beta ^2} $ is a relativistic factor of
a particle; $ \beta = v/c $, relative velocity of particles. Therefore
the hardness of the nonresonance radiation scattered by an electron is
proportional to $4\gamma ^2 $. At the same time ions are resonant
systems. They are excited by photons of a laser beam, and rescatter
them later. The nonmonochromatic laser beam is used in order to ions of
different energies scattered laser photons independently on their
energy. In this case the energy of laser photons interacting with ions
at resonance is inversely proportional to $2\gamma $. Therefore the
hardness of radiation scattered by these ions is proportional to $ 2
\gamma $. The difference in the dependence of hardness of radiation
scattered by electron and ion beams on their energy leads to the
difference 2 times in the expressions for equilibrium energy spreads,
longitudinal and transverse radial-phase dimensions of ion and electron
beams\footnote {Non-resonant crossection of scattering of laser photons
by ions is 10-15 orders less then the resonance one.}.

In this paper the detailed calculations of the equilibrium dimensions,
energy spread, the time of radiative damping of amplitudes of betatron
and phase oscillations of the particle beams interacting with the laser
beams in storage rings are presented. General formulas are received for
the non-stationary case, when the laser beam is displaced with some
velocity in the radial direction relative to the beam of particles and
such a way the degree of overlapping of beams is changed. The
calculations are carried out in the assumption, that the displacement
of the vertical position of the exited ions for their life-time can
be neglected. The received results include the possibility of the
description of muon beam dynamics when material targets are used in
storage rings (ionization cooling). The example is considered to
estimate the possible range of energies of electron storage rings and
hardness of laser and scattered radiation proceeding from necessary
life-time of the electron beam in the storage ring. Some applications
of the being cooled beams are discussed.

The parameters of the particle beams for stationary conditions of
interaction of laser and particle beams were derived in papers \cite
{idea, prl} for ion beams and in the paper \cite {zhirong} for electron
beams.  Dynamic conditions of interaction were investigated for the
case of switched off the radiofrequency system of the storage ring in
the paper \cite {ICFA01}.

       \section {Radiative effects in laser-electron and laser-ion
       storage rings in stationary conditions}

Below we will present in the general form a derivation of basic
equations describing the dynamics of interaction of particles with a
laser beam in storage rings and then we will discuss the losses of
particles caused by multiple scattering of the laser beam photons and
intrabeam scattering. The maximal stored current limited by coulomb
repulsion of particles in the beam will be discussed as well.

    \subsection {The equilibrium vertical dimension of a particle beam}

The dependence of a distance of a particle from the instantaneous orbit
of the storage rings in the vertical direction is described by the
equation

        \begin {equation} 
        z (s) = A _z (s) \cos \xi (s),
        \end {equation}
where $A _z (s) = C _ {z \, 0} \beta _z ^ {1/2} (s) $ is the amplitude
of particle oscillations; $ \xi (s) = \int _0 ^s ds/\beta _z (s) + \xi
_0 $, phase; $C _ {z \, 0} $, $ \xi _0 $, constants determined by the
initial conditions; $ \beta _z (s) $, $\beta$-function of the storage
ring; $s$, longitudinal coordinate \cite {kolom-leb} - \cite
{wiedemann}. The derivative

        \begin {equation} 
        z ^ {'} (s) = {A _z (s) \over \beta _z (s)} [{\beta _z ^ {'}
        (s) \over 2} \cos \xi (s) - \sin \xi (s)] = {\beta _z ^ {'} (s)
        \over 2 \beta _z (s)} z - {A _z (s) \over \beta _z (s)} \sin
        \xi (s).  \end {equation}

The square of the amplitude of vertical betatron oscillations,
according to (1), (2), can be expressed through the initial conditions
$z_0 $, $z _0 ^ {'} = d z/ds | _ {s=s_0} $ in some point $s _0 $

        \begin {equation} 
        A^2 _z (s) = {\beta _z (s) \over \beta _ {z \, 0}}
        A^2 _ {z \, 0} (s), \hskip 10mm A^2 _ {z \, 0} = z _0 ^ {2} +
        \beta _ {z \, 0} ^2 (z ^ {'} _0 - {\beta ^ {'} _ {z \, 0}
        \over 2 \beta _ {z \, 0}} z _0) ^2,
        \end {equation}
where $ \beta _ {z \, 0} $ is the value of the vertical
$\beta$-function in a point $s _0 $. Further we will take into account
the quantum nature of the interaction of particles with the laser beam.

At the moment of a photon scattering the coordinate of the particle is
not changed ($ \delta z =0 $). If a photon is emitted in a direction of
a particle velocity, then the momentum and inclination angle of the
particle trajectory is not changed as well ($ \delta z ^ {'} =0 $).
Therefore in the case of vertical oscillations the particle emission of
a photon leads to a change of its trajectory only in the case, when the
photon is emitted under some angle to its velocity.

If photon is scattered at some angle $ \theta \ne 0 $ to a vector of
velocity of a particle and at the scattering point $ \beta ^ {'} _ {z
\, sc} = 0 $, then the change of the square of the amplitude of
vertical betatron oscillations of the particle in a point of scattering

        \begin {equation} 
        \Delta A^2 _ {z \, sc} = \beta _ {z \, sc} ^2 [2z ^ {'} _ {sc}
        \delta z ^ {'} + (\delta z ^ {'}) ^2],
        \end {equation}
where

        \begin {equation} 
        \delta z _1 ^ {'} = {\hbar \omega \over \varepsilon _s} \sin
        \theta \cos \phi,
        \end {equation}
$ \hbar \omega $ is the energy of scattered photon; $\varepsilon _s $,
the equilibrium energy of the particle.

The rate of growth of a square of the amplitude of vertical betatron
oscillations of particles $A _{z \,1}$ caused by quantum processes of
scattering of laser photons is

        \begin {equation} 
        {d A ^2 _ {z \, 1} \over dt} = \int {\partial N ^2 _{\gamma}
        \over \partial o \partial t} \Delta A^2 _ {z \, sc} do,
        \end {equation}
where $\partial ^2 N _ {\gamma} /\partial o \partial t = f dN _ {\gamma}
/do $ is the angular distribution of the flow of scattered photons;
$f$, the frequency of the revolution of a particle at the orbit of the
storage ring; $do = \sin \theta d\theta d\phi $, the element of a
solid angle; $ \theta $, the angle between the laser beam axis
and the direction of observation; $ \phi $, the azimuth angle;
$dN _ {\gamma} /do = (1/\hbar \omega) (d\varepsilon ^ {rad} /do) $,
$d\varepsilon ^ {rad} / do = \varepsilon ^ {rad} F (\theta, \phi) $, $
\varepsilon ^ {rad} $, the energy of the particle being lost at the
passage of the laser beam (see section 2.4); $F (\theta, \phi)$,
the normalized angular distribution of the scattered radiation ($
\int F (\theta, \phi) do = 1 $). For circular and linearly polarized in
the plane of the particle orbit laser beam radiation in the conditions
of dipole radiation \cite {abb}

        $$ F (\theta, \phi) _ {cp} = {3\over 8 \pi \gamma ^4} [{1\over (1 -
        \beta \cos \theta) ^3} - {\sin ^2\theta \over 2\gamma ^2 (1 -
        \beta \cos \theta) ^5}], $$

        \begin {equation} 
        F (\theta, \phi) _ {lp} = {3\over 8 \pi \gamma ^4} [{1\over (1 -
        \beta \cos \theta) ^3} - {\sin ^2\theta \cos ^2 \phi \over
        \gamma ^2 (1 - \beta \cos \theta) ^5}]. \end {equation}

From equations (4), (5), (6) and the dependence of the frequency of
scattered radiation on the angle $ \omega = \omega _L (1 + \beta) / (1 -
\beta \cos \theta) $ follows a general expression for the rate of
growth of a square of amplitude of vertical betatron oscillations of a
particle caused by scattering of photons in an interaction region

        \begin {equation} 
        {d A ^ {2} _ {z \, 1} \over dt} = {\hbar \omega _L (1 + \beta) P
        ^ {rad} \over m ^2 c ^4 \gamma ^2} \beta _z (s) \beta _ {z \,
        sc} \int {\sin ^2 \theta cos ^2 \phi \over 1 - \beta \cos
        \theta} F (\theta, \phi) do,
        \end {equation}
where $P ^ {rad} = f \varepsilon ^ {rad} $ is the power of the
radiation scattered by the particle; $\beta _ {z \, sc} $, the vertical
$\beta$-function at the interaction region of the laser and particle
beams. It is supposed, that along the length of the interaction region
and along the length of a free path of ions in the exited state the
change of $\beta$-function and the displacement of the ion vertical
position can be neglected.  At the derivation of (8) we took into
account that at the averaging the first term in (4) tends to zero.

According to (7), (8) in the specific case of dipole scattering the
rate of growth of the square of amplitude of particle oscillations does
not depend on the kind of polarization of laser radiation and has
linear time dependence

        \begin {equation} 
        {d A ^ {2} _ {z \, 1} \over dt} = {6 \pi \hbar c P ^ {rad} _s \over
        5 \varepsilon _s ^2 \lambda _L} \beta _z (s) \beta _ {z \, sc}.
        \end {equation}

Within the framework of the classical electrodynamics the
electromagnetic radiation is emitted in the direction of the velocity of
a particle and it is not accompanied by a radiative damping of
amplitudes of vertical betatron of oscillations. The damping arises at
the acceleration of particles in the high-frequency resonator. In this
case particle changes longitudinal component of momentum only, and
consequently the angle of an inclination of a trajectory

         \begin {equation} 
         \delta z _2 ^ {'} = - z _ {sc} ^ {'} {q U _m \cos
         \varphi \over \varepsilon _s} \simeq - z ^ {'} _ {sc}
         {\varepsilon ^ {rad} _s \over \varepsilon _s},
         \end {equation}
where $q $ is the charge of a particle; $U _m $ and $ \varphi $,
amplitude and phase of the accelerating voltage at the moment of
passing by the particle of the resonator; $ \varepsilon ^ {rad} _s $,
the energy, emitted by equilibrium particle at one passing of the laser
beam. We took into account, that if the equilibrium energy of particles
in the storage rings is not changed, then the radiation losses of the
energy by a particle are equal to its energy increment in the
resonator. In this case in average $qU _m \overline {\cos \varphi} = qU
_m \cos \varphi _s = \varepsilon ^ {rad} _s $, where $ \varphi _s $ is
the equilibrium phase. The change of the square of amplitude of
vertical betatron oscillations of a particle in region of the location
of the resonator can be presented in the same form (4), if we will put
$ \beta _ {z \, rf} ^ {'} = 0 $, where now $ \beta _ {z \, rf} $ is the
value of the vertical $\beta$-function in this region.

From (4), (10) the expression for the rate of damping of the square of
the amplitude of vertical betatron oscillations of a particle caused by
the change of its energy in the resonator

        \begin {equation} 
        {d A ^ {2} _ {z \, 2} \over dt} = f \Delta A ^2 _z = - A _z ^2 {P _s
        ^ {rad} \over \varepsilon _s}. \end {equation}

We put $ \beta _z ^ {'} = 0 $ at the region of location of the
resonator, neglected the second term in (4), took into account, that $z
^ {'} _ {rf} = A _ {z \, rf} \sin \xi _n /\beta _ {z \, rf} $ and
the average value $ < {\sin ^2 \xi _n} > = 1/2 $, where $ \xi
_n $ is the value of the phase $ \xi $ at the revolution $n $.

The change of the square of amplitude of vertical betatron oscillations
of a particle caused both quantum processes of scattering of laser
photons and classical radiative damping simultaneously can be presented
in the form $ {d A^2 _ {z} / dt} = {d A ^ {2} _ {z \, 1} / dt} + {d A ^
{2} _ {z \, 2} / dt} $ or

        \begin {equation} 
        {d A^2 _ {z} \over dt} = - A_z ^2 {P ^ {rad} _s \over
        \varepsilon _s} + {6 \pi \hbar P ^ {rad} _s\over 5 \lambda _L
        \varepsilon _s^2} \beta _z (s) \beta _ {z \, sc}.
        \end {equation}

According to (12) the square of the equilibrium amplitude of vertical
betatron of oscillations of a particle is

        \begin {equation} 
        A^2 _ {z \, eq} = {d A ^ {2} _ {z \, 2} \over dt} {\tau _z \over 2} =
        {6 \pi \Lambda _c \over 5 \lambda _L \gamma _s}
        \beta _z (s) \beta _ {z \, sc},
        \end {equation}
and the time of radiative damping

            \begin {equation} 
            \tau _z = {2 \varepsilon _s \over P ^ {rad} _s},
            \end {equation}
where $ \Lambda _c = \hbar/ mc $ is the compton length of a wave of a
particle; $ \gamma _s = \varepsilon _s /mc ^2 $, the relativistic
factor of the equilibrium particles. For electron $ \Lambda _c \simeq
3.86 \cdot 10 ^ {-11} $ cm.

According to the central limiting theorem the density of distribution
of particles of the beam in a vertical plane is described by the Gauss
law

        \begin {equation} 
        \rho (z) = {1\over 2 \pi} \sigma _z e ^ {- {z^2\over 2\sigma _ {z}
        ^2}},
        \end {equation}
where $ \sigma _z = A _ {z \, eq} / \sqrt {2} $ is the dispersion of
the distribution of the density. Therefore the dispersion of the
density distribution and the emittance of the beam of particles $
\epsilon _z = \sigma ^2 _ {z \, eq} /\beta _z $ we can present in the
form

        \begin {equation} 
        \sigma _ {z} = \sqrt {3 \pi \Lambda _c \over 5 \lambda _L
        \gamma _s} \sqrt {\beta _z (s) \beta _ {z \, sc}}, \hskip 10mm
        \epsilon _ {z} = {3 \pi \Lambda _c \over 5 \lambda _L \gamma _s}
        \beta _ {z \, sc} \end {equation}

          \subsection {The equilibrium radial dimension of a beam
          determined by betatron oscillations}

The radial coordinate of a particle in a storage ring is $x = x _
{\eta} + x _b $, where $x _ {\eta} $ is the deviation of an
instantaneous orbit of this particle from equilibrium orbit and $x _b$,
the deviation of the particle from the instantaneous orbit. The
instantaneous orbit makes slow radial-phase oscillations relative to
the equilibrium orbit, and the particle makes fast betatron oscillation
relative to the instantaneous orbit.

The expression for the deviation of a particle from the instantaneous
orbit in the horizontal plane of the storage ring has the form
identical to the form for the vertical oscillations. It can be obtained
by replacing $z $ by $x _b $ in (1) - (3). Unlike vertical, the radial
betatron oscillations are connected with radial-phase oscillations.
The position of the instantaneous orbit of a particle is changed
intermittently after scattering of a laser photon by the value

        \begin {equation} 
        \delta x _ {\eta} = - {D _ {x \, sc} \over \beta ^2 _s} {\hbar
        \omega\over \varepsilon _ {s}},
        \end {equation}
where $D _x = p (\partial x _ {\eta} /\partial p) $ is the dispersion
function of the storage ring; $p = \beta \gamma $, relative momentum of
the particle. If $D _x \ne 0 $, then with the intermittent change of an
instantaneous orbit the amplitude betatron of oscillations will be
changed intermittently as well. At that the inclination of the particle
trajectory will not be changed.

The change of the square of the amplitude of horizontal betatron
oscillations of particles in the scattering point under the condition
that at this point $ \beta ^ {'} _ {z \, sc} = 0 $, $ \delta x _ {b} =
- \delta x _ {\eta} $, looks like

        \begin {equation} 
        \Delta A^2 _ {x \, b \, sc} = - 2 x _ {b \, sc} \delta x _
        {\eta} + (\delta x _ {\eta}) ^2 + 2\beta _ {x \, sc} ^2 x ^ {'}
        _ {b \, sc} \delta x _b ^ {'} + \beta _ {x \, sc} ^2 (\delta x
        _b ^ {'}) ^2.
        \end {equation}

The included in (18) values $ \delta x _b ^ {'} $ are similar to the
corresponding values (5), (10) for vertical oscillations:

        \begin {equation} 
        \delta x _ {b \, 1} ^ {'} = {\hbar \omega \over mc^2 \gamma}
        \sin \theta \cos \phi, \hskip 10mm \delta x _ {b \, 2} ^ {'} =
        - x _ { b \, 0} ^ {'} {q U _m \cos \varphi \over mc ^2 \gamma}.
        \end {equation}

In the general case $D _x $ depends on the longitudinal coordinate and
can accept zero values in some points and on the lengths of straight
sections of the storage ring. In the smooth approximation  $D _x \simeq
\alpha \overline R \simeq \overline R /\nu _x ^ {-2} $, where $ \alpha
= \partial \ln C/ \partial \ln p $, is the momentum compaction factor;
$C $ the perimeter of the trajectory; $ \nu _x$, the frequency of
radial betatron of oscillations; $ \overline R = C/2 \pi $, the average
radius of the storage ring. If $D _x = 0 $ then the  radial betatron
oscillations will be described by the expressions like (16) for the
vertical one if we will replace indexes $z $ by $x _b $:

        \begin {equation} 
        \sigma _ {x \, b} ^ {'} = \sqrt {3 \pi \Lambda _c \over 5
\lambda _L \gamma _s} \sqrt {\beta _x (s) \beta _ {x \, sc}}, \hskip
        10mm \epsilon _ {x \,} = {3 \pi \Lambda _c \over 5 \lambda _L
        \gamma _s} \beta _ {x \, sc}.
        \end {equation}

If $D _x \ne 0 $, then in the expression (18) for the square of
amplitude of oscillations, as a rule, can be neglected the
fourth item. In this case the rate of growth of the square of the
amplitude of radial betatron oscillations of particles caused by
quantum processes of scattering of laser photons determined, according
to (17), (19) by the values $ \delta x _ {\eta} $, $ \delta x _ {b \, 2}
^ {'} $, can be presented in the form

        \begin {equation} 
        {d A ^2 _ {x \, 1} \over dt} = \int {\partial N _ {\gamma} \over
        \partial \omega \partial t} \Delta A^2 _ {x \, b \, sc} d \omega
        = {7 D _x ^2 \hbar \omega _m P _ s ^ {rad} \over 10 \beta ^4 _s
        \varepsilon _s ^2} + {2 D _ {x \, sc} P ^ {rad} \over \beta _s ^2
        \varepsilon _s} x _ {b},
        \end {equation}
where $ {\partial N _ {\gamma} /\partial \omega \partial t} = (f/ \hbar
\omega) (\partial \epsilon _ {\gamma} /\partial \omega) $, $ \partial
\epsilon _ {\gamma} /\partial \omega = 3 \varepsilon ^ {rad} [1 - 2
\omega /\omega _m + 2 (\omega /\omega _m) ^2] \omega /\omega _m ^2 $ is
the spectral distribution of the energy of the scattered laser
radiation; $ \omega _m = (1 + \beta) ^2 \gamma ^2 \omega _L $, the
maximal frequency in the spectrum of the scattered radiation.

The rate of damping of the square of the radial amplitude of the
particle betatron oscillations caused by the change of its energy in
the radiofrequency resonator, according to (18), for the value $ \delta x
_ {b \, 2} ^ {'} $, determined by (19), can be presented in the form

        \begin {equation} 
        {d A ^ {2} _ {x \, 2} \over dt} = f \Delta A ^2 _ {x \, 2} = - A _ {x
        \, 2} ^2 {P _s ^ {rad} \over \varepsilon _s}.
        \end {equation}

At the derivation of (22) we put $ \beta _ {x \, rf} ^ {'} = 0 $ in
the region of location of the resonator and took into account, that
$x ^ {'} _ {rf} = A _ {x \, rf} \sin \xi _n /\beta _ {x \, rf} $, $ <
{\sin ^2 \xi _n} > = 1/2 $.

The change of the square of amplitude of radial betatron oscillations
of a particle caused by a quantum processes of scattering of laser
photons and classical radiative damping simultaneously can be presented
in the form $ {d A^2 _ {x} / dt} = {d A ^ {2} _ {x \, 1} / dt} + {d A ^
{2} _ {x \, 2} / dt} $ or

        \begin {equation} 
        {d A ^2 _ {x} \over dt} = - A _x ^2 {P ^ {rad} _s \over
        \varepsilon _s} + {7 \pi \hbar c (1 + \beta _s) ^2 \gamma _s
        ^2 P ^ {rad} _s D _ {x \, sc} ^2 \over 10 \lambda _L \beta ^4 _s
        \varepsilon _s^2} + {2 D _ {x \, sc} P ^ {rad} \over \beta _s ^2
        \varepsilon _s} x _ {b}.
        \end {equation}

The radiative damping time of the amplitude of radial betatron
oscillations of a particle according to (23)

            \begin {equation} 
            \tau _ {x \, b} = \tau _ {z} = {2 \varepsilon _s \over
            P ^ {rad} _s},
            \end {equation}
the square of equilibrium amplitude of these oscillations

        \begin {equation} 
        A^2 _ {x \, eq} = {0.7 \pi \Lambda _c (1 + \beta _s) ^2
        \gamma _s D ^2 _ {x \, sc} \over \beta ^4 _s \lambda _L},
        \end {equation}
the dispersion of the radial distribution of the density and emittance
of the particle beam

        \begin{equation}   
        \sigma _{x \,b } ^{"} = D _{x \, sc} \sqrt {0.35 \pi \Lambda _c
        (1 + \beta _s)^2 \gamma _s \over \lambda _L \beta _s ^4} \sqrt
        {\beta _x (s) \over \beta _{x \, sc}}, \hskip 10mm \epsilon
        _{x} = {0.35 \pi \Lambda _c (1 + \beta _s)^2 \gamma _s D ^2 _{x
        \, sc} \over \lambda _L \beta _{x\, sc}}.
        \end{equation}

At the derivation of (25) we have assumed, that the laser beam has
homogeneous density. Therefore $P ^ {rad} $ does not depend on $x _ {b}
$, on the average $ \overline x _ {b} = 0 $ and consequently the last
term in (23) can be neglected. According to (20), (26) the ratio $
\sigma _ {x \, b} ^ {"} / \sigma _ {x \, b} ^ {'} = \sqrt
{7/3} (1 + \beta _s) \gamma _s D _ {x \, sc} /\beta _ {x \, sc} \beta
_s ^2 $.

       \subsection {The equilibrium energy spread of a particle beam}

The energy of a particle in the storage rings $ \varepsilon $ produce
the synchrotron oscillations about the equilibrium energy $ \varepsilon
_s $.  The difference between the energy of a nonequilibrium particle
and the equilibrium energy is connected with the derivative
of the particle phase $d \varphi /dt = \dot \varphi = h (\omega _s -
\omega) $ by the dependence

         \begin {equation} 
         \Delta \varepsilon =
        {\varepsilon _s \over hK \omega _s} \dot \varphi,
        \end {equation}
where $K = - \partial \ln \omega _r/ \partial \ln \varepsilon = (\alpha
\gamma _s ^2 - 1) / (\gamma _s ^2 - 1) $, $ \omega _r = 2 \pi f $, $h $
is the harmonic order of the radiofrequency voltage \cite {kolom-leb}
- \cite {wiedemann}.

The equation of phase oscillations of a particle accelerated in the
resonator and losing the energy by scattering of the laser beam photons
can be received from the balance of the energy received by a particle
at one revolution $ (d \varepsilon /dt) T = q U _m \cos \varphi -
\varepsilon ^ {rad} $. Subtracting from the equation for the energy
balance of a nonequilibrium particle the same equation for the
equilibrium particle ($d \varepsilon _s/dt = 0 $, $q U _m \cos \varphi
_s = \varepsilon _s ^ {rad} $) and taking into account (27) we will
find

        \begin {equation} 
        \ddot \varphi + {h\omega _s ^2 K (\varepsilon ^ {rad} -
        \varepsilon ^ {rad} _s) \over 2 \pi \varepsilon _s} -
        {hq\omega ^2 _s K U _m \over 2 \pi \varepsilon _s} (\cos
        \varphi - \cos \varphi _s) = 0. \end {equation}

The linear equation of small phase oscillations, according to (28), has
the form

        \begin {equation} 
        \ddot \psi + {h\omega _s ^2 K (\varepsilon ^ {rad} -
        \varepsilon ^ {rad} _s) \over 2 \pi \varepsilon _s} +
        \Omega ^2 \psi = 0,
        \end {equation}
where $ \psi = \varphi - \varphi _s \ll 1 $, $ \Omega = \omega _s \sqrt
{ qhKU _m \sin \phi _s/ 2 \pi \varepsilon _s} $.

If the laser beam is both homogeneous and motionless, its transverse
dimensions are much greater than the transverse dimensions of the beam
of particles, then in the relativistic case the value $ \varepsilon ^
{rad} = \varepsilon ^ {rad \, s} (\gamma /\gamma _s) ^2 $ for the
backward compton scattering and $ \varepsilon ^ {rad} \sim D/1 + D)
$ for the backward rayleigh scattering in the field of the broadband
laser beam, where $D \sim \gamma $ is the saturation parameter \cite
{idea, prl}. In this case $ \varepsilon ^ {rad} - \varepsilon ^ {rad}
_s = (d \varepsilon ^ {rad} /d \varepsilon) (\Delta \varepsilon) = k _i
\varepsilon ^ {rad} _s (\Delta \varepsilon/ \varepsilon _s) $, where $k
_i = 2 $ for the backward compton scattering and $k _i = 1 / (1 + D) $
for backward rayleigh scattering. In the considered case, the equation
of the small phase oscillations, according to (27), (29), has a form

        \begin {equation} 
        \ddot \psi + {k _i P ^ {rad} _s \over \varepsilon _s} \dot
        \psi + \Omega ^2 \psi = 0.
        \end {equation}

The equation (30) has the solution of the form $ \psi = \psi _ {m} (t)
\cos \Omega ^ {'} (t - t _0) $, where $ \psi _m = \psi _ {m \, 0} \exp
(-t/\tau _s) $ is the amplitude of small phase oscillations,

            \begin {equation} 
            \tau _s = {2 \varepsilon _s \over k _i P ^ {rad} _s},
            \end {equation}
the damping time of the  synchrotron oscillations of the phase and
the energy of the particle; $ \psi _ {m \, 0} $, and $t _0 $, the
initial amplitude of phase oscillations and initial time; $ \Omega ^
{'} = \sqrt {\Omega ^2 - \tau _s ^ {-2}} $, the frequency of small
phase oscillations of the particle.

Usually $ \Omega \tau _s \ll 1 $, $ \Omega ^ {'} \simeq \Omega $. In
this approximation the square of amplitude $ \psi _ {m \, 0} $
expressed through the initial phase $ \psi _0 $ and initial derivative
of a phase $ \dot \psi _0 = (d\psi /dt) | _ {t=t _0} $ has a form $
\psi _ {m \, 0} ^2 \simeq \psi _0 ^2 + \Omega ^ {-2} \dot \psi _0 ^2 $.
Since the phase of the particle is not changed at the scattering of the
laser photons, then the change of the square of the amplitude of phase
oscillations after the emission of the photon of the energy $\hbar
\omega $ can be presented in the form

        \begin {equation} 
        \delta \psi _m ^2 = [2 \dot \psi _0 \delta \dot \psi + (\delta
        \dot \psi) ^2] \cdot \Omega ^ {-2},
        \end {equation}
where $ \delta \dot \psi = hK\omega _s (\hbar \omega /mc^2 \gamma _s) $.

The rate of growth of the average square of the amplitude of phase
oscillations of a particle caused by quantum processes of scattering of
laser photons can be presented in the form

        \begin {equation} 
        \overline {d \psi _m ^2 \over dt} = \int {\partial N _ {\gamma}
        \over \partial \omega \partial t} \delta \psi _m ^2 d \omega =
        {7\Lambda _c h ^2 K ^2 \omega _s ^2 \omega _m P ^ {rad} \over
        10 m c^3 \gamma _s ^2 \Omega ^2},
        \end {equation}

In the process of calculation of (33) we have neglected the first term
in $ \delta \psi _m ^2 $, since at the further integration over time
it will disappear. The rate of growth of the average square of the
amplitude of the energy oscillations of a particle, according to
(27), $d (\Delta \varepsilon) ^2 _m /dt $ $ = (\varepsilon _s ^2 \Omega
^2/h ^2 K ^2 \omega _s ^2) (d \overline {\psi _m ^2} /dt) $.

According to general rules (see (13)) the equilibrium values of the
average squares of the amplitudes of the transverse and longitudinal
oscillations of particles, and also amplitudes of oscillations of the
square of their energy are determined by the product of the rate of
growth of the appropriate values and the half of the time of radiative
damping. Therefore, the square of the amplitude of the phase
oscillations of the particle and the square of its amplitude of the
energy oscillations, according to (31), (33), are equal

        \begin {equation} 
        \overline {\psi _m ^2} = {11.2 \pi ^2 hK \Lambda _c \gamma _s
        \varepsilon _s \over q U _m \sin \phi _s \lambda _L k _i},
        \hskip 10mm \overline {(\Delta \varepsilon) ^2 _m} = {5.6 \pi
        \Lambda _c \gamma _s \varepsilon _s ^2 \over \lambda _L k _i}.
        \end {equation}

The dispersions of the particle distribution in the beam for the
radial-phase oscillations $ \sigma _ {x \, \eta} = D _x \beta _s ^ {-2}
$ $ (\sigma _ {\varepsilon} / {\varepsilon _s}) $, longitudinal
oscillations $ \sigma _ {l} = \psi _m \overline R / h \sqrt 2 $, and
for the energy spread $ \sigma _ {\varepsilon} / \varepsilon _s $ are
equal:

        \begin {equation} 
        \sigma _ {x \, \eta} = {D _ {x} \over \beta _s ^ {2}} \sqrt {2.8
        \pi \Lambda _c \gamma _s \over \lambda _L k _i},
        \hskip 10mm \sigma _ {l} = {K \overline R
        \omega _s \over \Omega} ({\sigma ^ {\varepsilon} \over
        \varepsilon _s}),
        \hskip 10mm {\sigma _ {\varepsilon} \over
        \varepsilon _s} = \sqrt {2.8 \pi \Lambda _c \gamma _s \over
        \lambda _L k _i}.
        \end {equation}

According to (26), (35) the ratio $ \sigma _ {x \, b} ^ {"} /\sigma _
{x \, \eta } = (1 + \beta _s) \sqrt {k _i D _ {x \, sc} \beta _ {x}
(s)} /2 \sqrt {2 D _ {x} \beta _ {x \, sc}} $.

\subsection {The power of radiation scattered by a particle}

The energy of the laser radiation scattered by a particle at the
passage of the laser beam depends on the distribution of the density
of the number of photons in the beam. For Gaussian beam the density is
distributed by the law

        \begin {equation} 
        n _L = {N _L \over 2\pi \sigma _L ^2 l _L} ({\sigma _ {L \, 0}
        \over \sigma _L}) ^2 e ^ {- {r ^2 \over 2 \sigma _ {L} ^2}},
        \end {equation}
where $N _L = \varepsilon _L /\hbar \omega _L $ is the number of
photons in the laser beam; $ \varepsilon _L $, the energy of the laser
beam; $l _L $, the length of the laser beam, $ \sigma _L = \sigma _ {L
\, 0} \sqrt {1 + s^2/ l _R^2} $, the dispersion of the distribution of
the laser beam in any point $s $ in the longitudinal direction; $ \sigma
_ {L \, 0} $, the dispersion in the point $s = 0 $ corresponding to the
waist of the laser beam; $l _R = 4 \pi \sigma _L ^2 / \lambda _L$,
rayleigh length\footnote {Usually in the laser technology the dimension
of the laser beam $w _L = 2 \sigma _L$ is introduced}.

If the longitudinal dimension of a particle beam $ \sigma _b \ll 4 l_R
$, then the change of the transverse dimension of the laser beam for
the time of its interaction with the particle beam can be neglected. In
this case the energy being lost by the particle moving at distances $r
\ll \sigma _L $ from the axis of the counterpropagated laser beam

        \begin {equation} 
        \varepsilon ^ {rad} = \hbar \overline \omega n _L (0) l _L
        \sigma | _ {\gamma \gg 1} = {\sigma \varepsilon _L \gamma ^2 \over
        \pi \sigma _ {L \, 0} ^2},
        \end {equation}
where $n _L (0) = n _L (r $ = $ s $ = $ 0) $, $ \sigma = 8 \pi r _p ^2
/3 $ is the crossection of scattering of laser photons; $r _p = q ^2/mc
^2 $, classical radius of the particle; $ \hbar \overline {\omega} =
\hbar \omega _ {max} /2 = 2 \gamma ^2 \hbar \omega _L $, average
energy of scattered photons. In the case of compton (thompson)
scattering of photons by electrons $ \sigma = \sigma _T \simeq 6.65
\cdot 10 ^ {-25} $ cm$ ^ 2 $ ($r _e \simeq 2.82 \cdot 10 ^ {-13} $ cm).
For the case of the rayleigh scattering of photons by ions the
crossection is 10-15 orders higher.

If bunches of the particle beams have the gaussian distribution, then
the flow of photons emitted by one bunch $ \dot N _ {\gamma} = \int n
_e l _p n _L l _L f \sigma dx dz = N _e N _L f \sigma /S _ {eff} $,
where $S _ {eff} = 2\pi ( \sigma _ {L \, 0} ^2 + \sigma ^2) $, $l _p $
is the length of the bunch; $N _p $, the number of particles in the
bunch. Therefore the power of radiation scattered by one bunch

        \begin {equation} 
        P ^ {rad} = \hbar \overline \omega \dot N _ {\gamma} | _ {\gamma
        \gg 1} = 2 \gamma ^2 \hbar \omega _L \dot N _ {\gamma} =
        {2 \gamma ^2 N _p P _L \sigma \over S _ {eff}},
        \end {equation}
where $P _L = f \varepsilon _L $ is the average reactive power of the
laser beam stored in the open resonator.

            \subsection {Influence of synchrotron radiation
            on dynamics of particles in storage rings}

Below, for a comparison, we will present the equilibrium parameters of
beams in storage rings determined by the influence of synchrotron
radiation on dynamics of particles in storage rings \cite {kolom-leb} -
\cite {wiedemann}.

If the losses are defined by the synchrotron radiation emitted in the
bending magnets of the storage ring, then the power of radiation and
the time of the radiative damping of amplitudes of vertical betatron
oscillations are equal accordingly

            \begin {equation} 
            \overline P _ {sr} = {2cq ^2 \gamma _s ^4\over
            3R\overline R}, \hskip 10mm
            \tau _z ^ {sr} = {3R\overline R\over cr_e \gamma _s ^3}
            \simeq 354.86 {R\overline R \over \gamma _s ^3} ᥪ
            \end {equation}
where $R $ is the instantaneous radius of an orbit of a particle in
bending magnets of the storage ring.

The damping times  $ \tau _x $, $ \tau _ {\varphi} $ of the horizontal
and longitudinal dimensions of a beam are approximately equal to $ \tau
_z $. At that $ \tau _x ^ {-1} + \tau _ {\varphi} ^ {-1} = 3\tau _z ^
{-1} $. The equilibrium transverse emittance and the dispersion of the
energy spread of the relativistic beam of particles are equal

            \begin {equation} 
            \epsilon ^ {sr} _ {z} \simeq {13 \sqrt 3 \over
            96} {\Lambda _c \over \nu _ {z}},
            \hskip 2cm \epsilon ^ {sr} _ {x} \simeq {C _ {\gamma} \Lambda
            _c \gamma _s ^2 \over \nu _ {x} ^3 J _x}, \hskip 2cm
            \left ({\sigma _ {\varepsilon} \over \varepsilon} \right)
            ^ {c¨} = \gamma _s \sqrt {C _ {\gamma} \Lambda _c
            \over 2 R J_s},
            \end {equation}
where $C_\gamma = (55\sqrt 3/ 96) \simeq 0.992 $ and $ J _ {x, s} \sim
1 \div 2 $ are constants determined by the magnetic structure of the
storage rings ($J _x + J _ {s} = 3 $).

The dispersion of distribution of the electron beam density determined
by the betatron and radial-phase oscillations in the smooth
approximation are equal

            \begin {equation} 
            \sigma ^ {sr} _ {z} = {1\over \nu _z} \sqrt {{13 \sqrt 3 \over
            96} \overline R \Lambda _c}, \hskip 10 mm \sigma ^ {sr} _ {x} =
            {\gamma _s \overline R \over \nu _ {x} ^2} \sqrt
            {C _ {\gamma} \Lambda _c \over R J _x},
            \hskip 10mm \sigma ^ {sr} _ {àä} = {\gamma _s \over \nu _ {x}
            ^2} \sqrt {C _ {\gamma} \Lambda _c R\over J _s}.
            \end {equation}

The dispersions of the density of distribution of an electron beam
over phases and in the longitudinal direction, according to (27),
are connected with the dispersion of distribution of particles
over the energy by the expressions

            \begin {equation} 
            \sigma _ {\eta} = {\hbar K \omega _s \over \Omega}
            ({\sigma ^ {\varepsilon} \over \varepsilon _s}),
            \hskip 10mm
            \sigma _ {l} = {K \overline R \omega _s \over \Omega}
            ({\sigma ^ {\varepsilon} \over \varepsilon _s}).
            \end {equation}

         \subsection {Life-time of particles in the storage rings}

The life-time of particles in the storage rings is limited by losses
caused by various processes. For the case of multiple scattering of
particles by photons of a laser beam

            \begin {equation} 
            \tau _ {loss} ^L = {e ^r \over 2 r} \tau,
            \end {equation}
where $r = A^2 _ {lim} /A ^2 _ {eq} $, $A ^2 _ {lim} $ and $A ^2 _ {eq}
= 2 \sigma ^2 $ is the square of the limiting amplitude and the square
of the equilibrium amplitude of oscillations of the particle
accordingly; $ \tau $, the time of the radiative damping \cite {bruk}.
In the case of phase oscillations $A ^2 _ {eq} = 2 \sigma _
{\varepsilon} ^2 $, $A ^2 _ {lim} = (\Delta \varepsilon) _ {sep} ^2 $,
where $ (\Delta \varepsilon) _ {sep} $ is the maximal deviation of the
energy of a particle moving along the separatrix from the equilibrium
energy

            \begin {equation} 
            (\Delta \varepsilon) _ {sep} = \sqrt {2 q V \varepsilon _s
            (\sin \varphi _s - \varphi _s \cos \varphi _s) \over \pi h
            K}.
            \end {equation}

In order to neglect the losses caused by multiple processes of quantum
fluctuation of synchrotron radiation it is necessary to fulfil the
condition $ (\Delta \varepsilon) _ {sep} \gg \sigma _ {\varepsilon} $.

The intrabeam interaction of particles of a beam results in loss them
from the region of stable phase oscillations (Touschek effect). The time
of life of particles in storage rings in the relativistic case

            \begin {equation} 
            \tau _ {loss} ^T = {8 \pi \gamma ^2 \sigma _z \sigma y
            \sigma _l \over c r _p ^2 N _p D (\xi)} [{(\Delta
            \varepsilon) _ {᥯} \over \varepsilon _s}] ^3,
            \end {equation}
where $D (\xi) = \sqrt \xi [- (3/2) \exp (-\xi) + (\xi/2) \int _ {\xi}
^ {\infty} (\ln u /u) \exp (-u) du + (1/2) (3 \xi - \xi \ln \xi +2)
\int ( \exp (-u) /u) du] $, $ \xi = [(\Delta \varepsilon) _ {᥯} \beta
_x/ \gamma \varepsilon _s \sigma _ {x}] ^2 $.

         \subsection {Limiting current of the storage rings }

The limiting number of particles in the beam of the storage ring is
determined by the volume charge of the beam (coulomb repulsion). In the
relativistic case \cite {wiedemann}

         \begin {equation} 
         N _ {lim} = {\sqrt {2 \pi} \nu _z \Delta \nu _z \gamma ^3
         \over \overline R r _e} {\sigma _l \over \overline R}
         \sigma _z (\sigma _z + \sigma _x).
         \end {equation}

                     \subsection {Example}

In the Tables 1 - 3, as an example, the basic parameters of the
electron storage ring, the electron and laser beam parameters and the
parameters of the beam of scattered radiation are presented. It is
supposed, that the interaction of the electron and circular polarized
laser beams occurs in the straight section of the storage ring having
zero dispersion function and small $\beta$-function ($ \beta _ {x, z /,
sc} = 1$ sm).  Outside of the region of interaction $ \beta _ {x, z}
(s) \simeq \overline R/\nu _ {x, z} = 10 $ cm, $D _x = \overline R/ \nu
^2 = 1 cm $. The storage ring has large energy acceptance $ (\Delta
\varepsilon) _ {A} / \varepsilon _ {s} = \pm 0.1 $. One electron bunch
regime is used.

From the tables we can see, that in the conditions of small currents
horizontal emittance and the horizontal dimensions of the electron beam
determined by the synchrotron radiation (the laser is switched off) are
much greater then the corresponding dimensions determined by backward
compton scattering, if the interaction of the electron and laser beams
occurs in the region with zero dispersion function. The vertical
dimension of the beam determined by the synchrotron radiation is much
less then the horizontal one. The vertical and horizontal dimensions
of the beam determined by backward compton scattering are identical.
The increase of the dispersion function of the storage ring to the
value $D _x \simeq \overline R / \nu ^2 $ leads to the increase of the
horizontal dimension of the beam by the backward compton scattering
approximately $ \gamma _s = 10 ^2 $ times.

The conditions of small currents in this example correspond to the
currents $ \sim 0.1 $ mA. At the current 100 mA we are forced to
increase the dimensions of a beam artificially so that outside of the
regions of interaction they were equal to the dimensions $ \sigma _z
\cdot \sigma _x \cdot \sigma _l = 75 \cdot 150 \cdot 4000 \mu$m. At
that the life-time of a beam caused by intrabeam scattering of
electrons (Touschek effect) reaches the acceptable values $ \tau _ {loss}
^ {T} \simeq 10 ^3 $ ᥪ, and the coulomb repulsion shifts the
frequencies of the betatron oscillations to the value $ \Delta \nu
\simeq 0.2 $, that is close to the extremely possible shift\footnote
{On practice the shift $\Delta \nu \simeq 0.4 $ is achieved \cite
{wiedemann}.}. The increase of the horizontal dimension of a beam can
be achieved by a selection of a final value of the dispersion function
at the interaction point (IP) or by excitation of betatron and phase
oscillations of particles of the beam by a high-frequency fields. The
vertical dimension of a beam can be increased by coupling of vertical
and horizontal oscillations.  Besides the coulomb repulsion the
intrabeam scattering of the electrons leads to the increase of the
transverse dimensions of the beam in this case \cite {wiedemann}.

The equilibrium energy spread of a beam of particles does not depend on
the focusing properties and parameters of the radio-frequency system of
the storage ring.  It grows with the increase of the energy, decreases
with increase of the laser wavelength ($ \sim \sqrt {\gamma _s/\lambda
_L} $) and at the energy 50 MeV has rather large value ($ \sigma _
{\varepsilon} /\varepsilon _s = 1.3 \cdot 10 ^ {-2} $), and leads to
hard requirements to the energy acceptance of the storage ring, and,
hence, to the focusing forces. In this example we have limited the
acceptance of the storage ring by the value $ (\Delta \varepsilon)
_A / \varepsilon _ {s} = \pm 0.1$\footnote {There are projects of
storage rings, in which the energy acceptance is equal $ (\Delta
\varepsilon) _A/ \varepsilon _ {s} = \pm 0.25 $ \cite {klotz}.}.
Therefore with increase of the energy of particles in the storage rings
it is necessary to go to the more long wavelength lasers. At the energy
of particles 1 Ē', $ \lambda _L = 20 $ $ \mu$m the maximal energy of
scattered photons is increased from $ \varepsilon _ {\gamma \, m} =
\hbar \omega _m = 50 $ keV up to $ \varepsilon _ {\gamma \, m} = 1.0 $
MeV, and the energy spread of the beam of particles is not changed ($
\sim 1.3 \% $).

The limiting currents in the storage ring are increased proportionally
to the cube of the energy and in the inverse proportion to the electron
beam crossection in the storage ring. Therefore the increasing of the
energy of the storage ring will permit to reduce the dimensions of the
beams and to increase the stored current. The transverse dimensions of
the laser beam must be greater then the transverse dimensions of the
electron beam (in the considered example $ \sigma _L = 50 $ $ \mu$m, $
\sigma _ {x, z} = 23.7 $ $ \mu$m in the region of interaction). The
decreasing of the electron beam current up to 10 mA and the appropriate
decreasing of the transverse dimensions of the electron and laser beams
the power of scattered radiation is not changed, and the brightness is
increased.

The basic advantage of the storage rings compared with the linear
accelerators as the sources of the hard electromagnetic radiation
consists in the fact that the scattering of the laser photons by
electrons do not lead to losses of these electrons, and the electrons
are used repeatedly. In the considered case the losses of electrons are
determined by the multiple processes of scattering in the field of the
laser beam and by Touschek effect. For the life-time $ \tau _ {loss} ^T
\simeq 10 ^ {3} $ ᥪ the electron makes $N _ {eff} = f \tau _ {loss}
\simeq 10 ^ {11} $ revolutions and lose the energy $ \Delta \varepsilon
_ {loss} = e V _s N _ {eff} \simeq 10 ^ {13} $ eV $ \sim 10 ^7
\varepsilon _s $.  Therefore without losing of the efficiency the
life-time of the beam can be decreased, and the frequency of the
injection of electrons can be increased. Thus the requirements to
vacuum in the storage ring and to the value of the energy acceptance
can be reduced.

The energy of the electron is changed by the value $ \varepsilon _
{\gamma} \ll \varepsilon _ {sep} < (\Delta \varepsilon) _ {A} $ in the
process of scattering of the laser photons. Therefore the
single losses of electrons in this process are absent.

It is interesting to note, that in the considered case the average
value of the jump of the electron energy caused by scattering a photon
is much greater then the increase of the energy of the equilibrium
electron per one revolution ($ {\varepsilon _ {\gamma \, m}} /2 \gg eV
_s $), and the probability to the electron to scatter a photon at the
passage of the laser beam is small ($ 2 e V_s / (\Delta \varepsilon) _
{\gamma \, m} \simeq 4.0 \cdot 10 ^ {-2} $). It means, that the
trajectory of the electron produce on average two jumps at one period
of phase oscillations ($ \Omega / \omega _s = 2 \cdot 10 ^ {-2} $).
At the same time the above mentioned expressions for the equilibrium
amplitudes of phase oscillations remain valid, as at their derivation
we are not used any restrictions on probability of the emitting of
photons.

In the considered storage ring, as well as in the similar storage ring
discussed in \cite {zhirong}, the harmonic order of the radio-frequency
generator is chosen equal to $h = 60 $. It corresponds to the
wavelength of the generator $ \lambda _ {rf} \simeq 10.5 $ cm. Such
harmonic order is convenient at the construction of the
small-dimensional radio-frequency resonator. However it is possible to
reduce the harmonic order of the generator, the voltage and the quality
of the resonator. So at the harmonic order $h = 4 $ ($ \lambda _ {rf} =
1.57 $ m) the amplitude of the voltage at the gap of the resonator will
be decreased to the value $V _m = 16.7 $ kV, and the dimension of the
separatrix will not be changed ($ (\Delta \varepsilon) _ {sep}
/\varepsilon _ {s} \simeq 0.073 $). At that the equilibrium transverse
dimensions of the beam will not be changed, and the longitudinal
dimension will be increased up to acceptable dimension $ \sigma _l ^
{bcs} = 7.4 $ mm $ < l _R $.  The equilibrium voltage in this case
remains much smaller then the amplitude of radio-frequency voltage. The
analogical remark can be done relatively to the focusing force. It is
important to have the zero dispersion function and low $\beta$-function
in the straight section of the storage ring. Less important is to have
very high tunes.

The use of supermirrors and storage rings with straight sections having
zero dispersion function allows to lower the beam dimensions and to
increase the power of the laser radiation in the resonator up to the
values, when the losses of the energy of the electron in the storage
ring through the backward compton scattering of the laser photons
become much greater than the losses of the energy through the
synchrotron radiation. Thus the radiative damping times of the
amplitudes of radial and phase oscillations are decreased essentially.
It allows to lower the influence of the intrabeam scattering inside the
electron beam and the influence of the emission of hard backward
scattered laser photons on the excitation of betatron and phase
oscillations.

The considered source has the emittance $ \epsilon _ {x, z \, b} ^
{bcs} = 7.3 \cdot 10 ^ {-9} $ cm in a conditions of small currents ($i
< 1 $ mA) and the emittance $ \epsilon _ {x, z \, b} ^ {bcs} = 5.4
\cdot 10 ^ {-5} $ cm at the current $i = 100 $ mA. It can be classified
as the light source of the third generation\footnote {According to
accepted classification \cite {winick} - \cite {hara} the light sources
of the third generation have emittances $ \epsilon _ {x, z \, b} ^
{bcs} < 1.0 \cdot 10 ^ {-6} $ cm, and the light sources of the fourth
generation will have the emittances $ \epsilon _ {x, z \, b} ^ {bcs} <
1.0 \cdot 10 ^ {-9} $ cm.}. The compactness, small cost, smooth tuning
of the frequency, narrow range of generated waves, high average photon
flux and brightness discovered in the papers \cite {zhirong}, \cite
{j.chen} are invaluable quality of such sources. They, without doubts,
can find the wide region of applications.

The storage ring RHIC, considered in \cite {prl}, \cite {epac96} allows
to receive ion beams $ ^ {129} _ {54} Xe ^ {47 +} $ with the energy
spread $ \left ({\sigma _ {\varepsilon} / \varepsilon} \right) ^ {bcs}
= 1.2 \cdot 10 ^ {-4} $ and the emittance $ \epsilon _ {x, z \, b} ^
{bcs} = 8.0 \cdot 10 ^ {-12} $ cm at the relative energy $ \gamma \sim
100 $. This is the more expensive and bulky storage ring. At the same
time it allows to proceed to the light sources of the fifth generation
\cite {esrf96}.

         \subsection {Some applications of the backward compton and
         backward rayleigh sources of hard electromagnetic radiation}

For the first time the sources of the backward compton scattering were
discussed in the papers \cite {lan}, \cite {lam}. Coherent scattering
of electromagnetic waves by electron beams was considered. Later the
incoherent scattering of photons of a laser beam by ultrarelativistic
electrons was considered in the papers \cite {aru}, \cite {mil} and
realized in papers \cite {fiocco} (nonrelativistic beams) and \cite
{kulik}, \cite {bem} (relativistic beams). The scattered $\gamma
$-quanta were used in the fields of nuclear physics and diagnostics of
electron beams.

Later the production of circular polarized backward compton $ \gamma
$-quanta and using them in the conversion systems for the production of
the longitudinally polarized positron beams and in nuclear physics was
discussed in papers \cite {mikh} - \cite {omori}\footnote {The scheme
using the helical undulator was discussed in these papers as well.} In
the paper \cite {mar} the intraresonator scheme of interaction of
electron and laser beams was considered. This scheme was realized later
in papers \cite {glotin}, \cite {litv} (in the regime of the
free-electron laser). In the paper \cite {tomi} the current 0.5€ was
stored in the storage ring NIJI-1 SR (maximum energy 270 MeV) at the
energy 163 MeV. This and similar papers stimulated the interest to
investigations of compact storage rings as the sources of x-ray
radiation for medical and other applications \cite {csonka}. In the
paper \cite {chen} the use of supermirrors was suggested to increase
the power of the scattered radiation.

In papers \cite {ispirian} - \cite {basov} the attention was paid to
the rayleigh scattering sources. In this case the scattering
crossection of a photon by bound electron in ion at resonance is many
orders ($\sim 10\div15$) higher then the compton one and hence the
power of scattered radiation is higher.

In the papers \cite {idea}, \cite {prl}, it was shown that the
scattering of laser photons by ions in the storage rings leads to the
three-dimensional radiative cooling of ion beams and, simultaneously,
to the excitation of the amplitudes of betatron and phase oscillations
of ions in the beam. The equilibrium parameters of ion beams according
to developed theory will differ from similar parameters of beams
determined by the synchrotron radiation. To reduce the influence of the
quantum excitation of amplitudes of betatron and radial-phase
oscillations it was suggested to use the zero dispersion function at
the IP\footnote {The using of damping wigglers and undulators in
storage rings, installed in the straight sections, with zero dispersion
function has allowed to reduce the emittances of electron beams (three
orders) \cite {wiedemann} and to proceed to light sources of the third
generation.}.

In the paper \cite {zhirong} it was shown, that the use of radiative
damping in electron storage rings of the intermediate energy,
supermirrors and modern lasers allows to generate the beams of
scattered photons with the power exceeding the power of the synchrotron
radiation from the orbit of these storage rings at higher hardness and
brightness. At that the damping time of amplitudes of electron
oscillations is decreased, the influence of the intrabeam scattering on
the emittance of the beam is decreased. This paper stimulated a series
of papers devoted to sources of backward compton scattering and their
applications \cite {glad} - \cite {urakawa}, \cite {ICFA01}. In papers
\cite {bes3} - \cite {clen} the attention was paid on the opportunity
of effective self-polarization of the positron beams in the storage
rings with special magnetic structure in the field of the circular
polarized laser beam.

       \section {Radiative effects in laser-electron and
       laser-ion storage rings in dynamics}

In the previous section we have considered the dynamics of particles in
storage rings in the field of a homogeneous counterpropagated laser
beam in stationary regime. In this case the transverse dimensions of
the laser beam were greater, than the dimensions of the particle beam
and their axes coincided. At the same time we have derived the general
formulas (23), (28) allowing to consider the dynamics of particle beams
interacting with an arbitrary targets (undulators, laser beams,
material media). The targets could be non-uniform and could be moved
(oscillate) in the radial direction (or, the orbits in the storage
rings could be moved in the backward direction). These formulas are

        \begin {equation} 
        {d A ^2 _ {x} \over dt} = - A _x ^2 {P ^ {rad} _s \over
        \varepsilon _s} + {7 \pi \hbar c (1 + \beta _s) ^2 \gamma _s
        ^2 P ^ {rad} _s D _ {x \, sc} ^2 \over 10 \lambda _L \beta ^4 _s
        \varepsilon _s^2} - {2 D _ {x \, sc} P ^ {rad} \over \beta _s ^2
        \varepsilon _s} x _ {b},
        \end {equation}

        \begin {equation} 
        \ddot \varphi + {h\omega _s ^2 K (\varepsilon ^ {rad} -
        \varepsilon ^ {rad} _s) \over 2 \pi \varepsilon _s} -
        {hq\omega ^2 _s K U _m \over 2 \pi \varepsilon _s} (\cos
        \varphi - \cos \varphi _s) = 0. \end {equation}

The values $ \varepsilon ^ {rad} $ and $P ^ {rad} = f\varepsilon ^
{rad} $ in (47), (48) usually depend on the energy of particles,
effective thickness of a target in a radial direction and on time by
the law $ \varepsilon ^ {rad} = F _1 (x, t) F _2 (\gamma) $, where $x =
x _b + x _ {\eta} $ is the deviation of the particle from the
equilibrium orbit. The deviation $x _ {\eta} $ is determined by the
deviation of the relative energy $ \gamma $ from the equilibrium one
and by the dispersion function .  For the stationary regime and
non-uniform target the value

     \begin {equation} 
     \varepsilon ^ {rad} = \varepsilon _s ^ {rad} + \varepsilon _s
     ^ {rad} {\partial \ln F _1 \over \partial x} (x _b + x _ {\eta}) +
     \varepsilon _s ^ {rad} {\partial \ln F _2 \over \partial \ln
     \gamma} {\Delta \gamma \over \gamma _s}.
     \end {equation}

In the case of cooling of the electron and ion beams in the field of
the laser beam the second term in (49) is equal $ \partial \ln F _2 /
\partial \ln \gamma = k _i $. In a case of ionization cooling of muon
beams this term is much less then unit and has more complicated
dependence on energy.

The second term in (49) is differ from zero only in the case, when the
target is non-uniform in radial direction (axis of a beam of particles
is on the slope near to caustic of the laser beam or, in case of
cooling of muon beams, wedge target is used).

The expressions (47) and (48), according to (49), can be presented in
the form

        \begin {equation} 
        {d A ^2 _ {x} \over dt} = - A _x ^2 {P ^ {rad} _s \over
        \varepsilon _s} (1 - {D _ {x \, sc} \over \beta _s ^2}
        {\partial \ln F _1\over \partial x}) + {7 \pi \hbar with (1 +
        \beta _s) ^2 \gamma _s ^2 P ^ {rad} _s D _ {x \, sc} ^2 \over
        10 \lambda _L \beta ^4 _s \varepsilon _s^2},
        \end {equation}

        \begin {equation} 
        \ddot \psi + {P ^ {rad} _s \over \varepsilon _s} (k _i +
        {D _ {x \, sc} \over \beta _s ^2} {\partial \ln F _1\over
        \partial x}) \dot \psi + \Omega ^2 \psi = 0.
        \end {equation}

According to (50), (51) the increase of damping decrement of amplitudes
of betatron oscillations of particles leads to the decrease of the
damping decrement of amplitudes of phase oscillations in the same
degree. The sum of decrements is a constant (analogue of the Robinson
criterion of damping). The decrements can accept negative values.

The non-stationary regime of interaction of being cooled particle beams
with targets was developed in \cite {ICFA01} for the case of the
unbunched beams. Switching on a radio-frequency accelerating field will
complicate the problem. This problem will be considered in the separate
paper.

            \section {Conclusion}

In this paper the detailed derivation of the equations describing the
dynamics and equilibrium dimensions of the particle beams interacting
with the laser beams in storage rings is presented. Brief derivation of
these equations was given in the papers \cite {idea, prl} for ion beams
and in the paper \cite {zhirong} for the electron beams\footnote {A
misprint is in the equilibrium energy spread $(\sigma _{\delta} =
\sigma _{\varepsilon}/\varepsilon)$ in \cite {zhirong} (corrected
in \cite {zhirong2}). In \cite {prl} the value $(\sigma _{\delta} = 2
\sigma _{\varepsilon}/\varepsilon)$}. The general non-stationary case
of interaction of particles with a target is considered. The example is
presented to illustrate the problem to estimate the possible range
of the energy of electron storage rings and the hardness of scattered
radiation proceeding from life-time of the electron beam in the ring.  Some applications of the being cooled beams were discussed.

The author would like to thank Prof. Mitsuru Uesaka, A.Mikhailichenko,
and Zhirong Huang for useful references and discussion of some topics
of this paper.

\newpage
         \addcontentsline {toc} {section} {\protect\numberline {6
         \hskip 2mm References}}

\newpage

\begin {table}
{ \large \hskip 4mm TABLE 1. \hskip 3 mm  Electron storage ring
parameters ($J _x =1, $ \hskip 2mm $ J _ {\varphi} =
2 $)} $ ^ * $. \\
\vskip 1mm
\begin {tabular} {|l|l |}
\hline Energy [MeV] & $ \varepsilon = 50 $ \\
\hline
Average radius [m] & $ \overline R = 1 $  \\
\hline
Instantaneous radius [m] & $R = 0.5 $  \\
\hline
Horizontal/vertical tunes & $ \nu _ {x, y} = 10 $ \\
\hline
$\beta$-function [cm] &$ \beta _ {x, z} = 10 $,
$ \beta _ {x, z \, sc} = 1 $ \\
\hline
No of electrons (aver. current [mA]) &$N _e = 1.3 \cdot
10 ^ {10}$ ($\overline i = 100 $)  \\
\hline
Current in a bunch [A] &$i _b = (C/2 \sigma _l) \overline i = 74$\\
\hline
Frequency of resonator [MHz] &
2856 ($\lambda _{rf} = 10.5 $ cm) \\
\hline
Harmonic order& $h = 60 $ \\
\hline
RF peak voltage [kV]  & $ 250 $ \\
\hline
Damping time of amplitudes &$ \tau _ {z} ^ {sr} =
1.774 $, \\
of radial betatron and &$ \tau _ {x, z} ^ {bcs} = 1.28 \cdot
10 ^ {-2} $  \\
phase oscillations [sec] &$ \tau _ {\varphi} ^ {bcs} = 6.5 \cdot
10 ^ {-3} $  \\
\hline
Equilibrium energy spread & $ \left ({\sigma _ {\varepsilon} /
\varepsilon} \right) ^ {sr} = 4.38 \cdot 10 ^ {-5} $ \\
  &$ \left ({\sigma _ {\varepsilon} /
\varepsilon} \right) ^ {bcs} = 1.3 \cdot 10 ^ {-2} $ \\
\hline
Equilibrium transverse emittance [cm]& $ \epsilon _x
^ {sr} = 7.7 \cdot 10 ^ {-7} $ \\
($D _x = 0 $, $ \beta _ {z, x \, sc} = 1 $ cm) &
$ \epsilon _ {x, z \,b} ^ {bcs} = 7.3 \cdot 10 ^ {-9} $ \\
\hline
Transverse dimensions at IP [$ \mu$m ] &$ \sigma _ {x \, b}
^ {sr} = 8.75 $ \\
($D _ {x \, sc} = 0 $, $ \beta _ {z, x \, sc} = 1 $ cm) &
$ \sigma _{x, z \,b} ^ {bcs} = 0.854 $  \\
\hline
Equilibrium transverse emittance [cm] &$ \epsilon _x ^ {sr}
= 7.7 \cdot 10 ^ {-7} $  \\
($D _{x \, sc} = 0.5 $ cm, $ \beta _ {z, x \, sc} = 1 $ cm) &
$ \epsilon _{x} ^{bcs} = 5.4 \cdot 10 ^ {-5} $  \\
\hline
Transv. beam dimensions at IP $ [\mu$m] &$ \sigma _ {x, z \, b}
^ {bcs} = 23.7 $  \\
($D _ {x \, sc} = 0.5cm $, $ \beta _ {z, x\, sc} = 1 $ cm) &  \\
\hline
Transv. beam dimensions  &$ \sigma _ {x \, b}
^ {bcs} = 75 $, $\sigma _ {x \, \eta} ^ {bcs} = 65 $, \\
outside of the IP [$ \mu$m] & $ \sigma _ {x} \simeq 10 ^2 $ \\
\hline
Longitudinal beam dimensions [mm] &$ \sigma _ {l}
^ {c¨} = 8.49 $ \\
 &$ \sigma _ {l} ^ {bcs} = 3.8 $ \\
\hline
Relative frequency & $ \Omega /\omega _s = 2.2 \cdot
10 ^ {-2} $ \\
\hline
Maximal energy deviation &$ (\Delta \varepsilon)
_ {sep} / \varepsilon _ {s} = 0.073 $ \\
\hline
Energy acceptance &$ (\Delta \varepsilon) _ {A} /
\varepsilon _ {s} = \pm 0.1 $ \\
\hline
Life-time [sec]  &$ \tau _ {loss} ^C = 5.6 \cdot 10^3 $, \\
 &$ \tau _ {loss} ^T = 7.1 \cdot 10 ^2 $  \\
\hline
Accepted beam dimensions at IP [$\mu$m] & $ \sigma _ {x \,b \, sc}
^ {bcs} = 23.7 $, $ \sigma _ {x \, \eta} ^ {bcs} = 0 $, \\
   &  $ \sigma _ {l} = $ $4 \cdot 10 ^3$ \\
\hline
Accepted beam dimensions outside IP  [$\mu$m]  &
$ \sigma _ {x \, b \, sc} ^{bcs} = 75 $, $ \sigma _
{x \, \eta} ^ {bcs} = 130 $,\\
$(\sigma _ {x \, sc} ^ {bcs} = \sqrt
{(\sigma _ {x \,b sc} ^ {bcs}) ^2 + (\sigma _ {x \, \eta}
^ {bcs}) ^2} $) & $ \sigma _{x \, sc} ^ {bcs} = 150 $,
$ \sigma _ {l} = 4 \cdot 10 ^3 $ \\
\hline
Limit. number of electrons for $ \sigma _z \cdot \sigma _x
\cdot \sigma _l$   & $ N _ {lim} = 1.34 \cdot 10 ^ {10} $ \\
$= 75 \cdot 150 \cdot 4200 $ $ \mu$m$^3$; \hskip 3mm $ \Delta
\nu = 0.2 $ & $ (\overline i _ {lim} = 100 $ mA) \\
\hline
\end {tabular}

\vskip 3mm
{ $ ^ * $ \small \hskip 1mm Emittances and dimensions of beams are
calculated in one-particle \\approximation (coulomb repulsion and
intrabeam scattering are neglected.)} \\
\end {table}

\begin {table}
{ \large \hskip 25mm TABLE 2. \hskip 5mm Laser and optical resonator
parameters.} \\
\vskip 1mm
\begin {tabular} {|l|l|}
\hline Wavelength [cm]& $ \lambda _L = 1 \cdot 10 ^ {-4} $ \\
\hline Photon energy [eV] & $ \hbar \omega _L = 1.24 $  \\
\hline Laser pulse length [cm]& $l _L = K\lambda _L =1.0 $ \\
\hline Distance between wave packets [m] & $L _ {wp} = 2 \pi
\overline R = 2 \pi \simeq 6.3 $ \\
\hline  Laser beam waist radius [cm] & $ \sigma _ {L} = 5 \cdot 10
^ {-3} $  \\
\hline
Rayleigh length [cm] & $l _ {R} = 4 \pi \sigma _ {L} ^2 /
\lambda _L = 3.14 $  \\
\hline
Laser bunch energy in resonator [J] & $ \varepsilon _L = 2 $ J \\
\hline
Resonator quality & $Q = 1 / (1 - R _r) = 1.0 \cdot 10 ^4 $ \\
\hline
Damping time of laser beam in resonator [sec] & $ \tau _r = Q L
_ {wp} /c = 2.1 \cdot 10 ^ {-4}  $ \\
\hline
Reactive laser pulse power in resonator [GW] &$P _b = 60 $ \\
\hline
Pulse intensity in the waist [$ (W/cm) ^2$ ] &$I = P _b/
2\pi \sigma _L ^2 = 3.82 \cdot 10 ^ {14}$  \\
\hline
Magnetic field strength in the waist & $ B _ {\perp m} =
\sqrt {2P _L/c \sigma _L ^2} \simeq 6.8 \cdot 10 ^5 $ Gs \\
\hline
Deflecting parameter & $p _ {\perp m} = e B _m/2 \pi m c^2 = 6.4 \cdot
10 ^ {-3}  $ \\
\hline
Electron energy losses per revolution [keV] & $e V _s = 1.06 $ \\
\hline
Laser pulse duration [sec] & $ \tau _L = 10 ^ {-3} $  \\
\hline
Frequency of laser pulses [Hz] & $f _L = 50 $ \\
\hline
Laser pulse power [kW] & $P _L = \varepsilon _L \cdot f /Q =
9.5 $ \\
\hline
Average laser power [W] & $ \overline P _L = P _L \cdot \tau _L
f _L = 475 $  \\
\hline
\end {tabular}
\end {table}

\vspace {5mm}
\begin {table}
{ \large \hskip 5mm TABLE 3. \hskip 6 mm Parameters of
the X-ray beam.} \\
\vskip 1mm \begin {tabular} {|l|l|}
\hline
The maximal energy of photons [keV] & $ \varepsilon
_ {\gamma \, m} = 50 $  \\
\hline
Pulse power of the beam [W] & $ P _ {imp} ^ {rad} = 95 $  \\
\hline
Average power [W] & $ \overline {P ^ {rad}} =
P _ {imp} ^ {rad} \cdot \tau _L \cdot f = 4.74 $  \\
\hline
Photon flux [sec$^{-1}$] & $ {\dot N _ {\gamma \, imp}} =
P ^ {rad} /\hbar \overline \omega = 2.37 \cdot 10 ^ {16}$\\
\hline
Photon flux in the range & $ \Delta {\dot N _ {\gamma \, imp}}
= (3 P ^ {rad} / \hbar \omega _m) (\Delta \omega /\omega _m)$\\
$ \Delta \omega /\omega _m = 0.1$ [sec$^{-1}]$ &
$ = 3.6 \cdot 10 ^ {15}$ \\
\hline
Average photon flux [sec$ ^{-1}]$ & $ \overline {\dot N _{\gamma}}
= \dot N _ {\gamma \, imp} \cdot \tau _L \cdot f = 1.19 \cdot
10 ^ {14} $  \\
\hline
Average photon flux in a range & $ \Delta \overline {\dot N _
{\gamma}} = {\dot N _ {\gamma \, imp}} \cdot \tau _L \cdot f$ \\
$ \Delta \omega /\omega _m = 0.1$ [sec$^{-1}]$ & $ = 1.8 \cdot
10 ^ {14} $  \\
\hline
Duration of pulses [sec] & $ \tau _ {x-ray} = 10 ^ {-3}$ \\
\hline
Frequency of pulses [Hz] & $f _ {x-ray} = 50 $  \\
\hline
Duty cycle & $S = \tau _ {x-ray} \cdot f _ {x-ray} = 5
\cdot 10 ^{-2}$ \\
\hline
\end {tabular}
\end {table}


\begin {thebibliography} {9}

\bibitem {idea} 
E.G.Bessonov, Journal of Russian Laser Research, 15, No 5, (1994),
p.403; Proc. of the Internat. Linear Accel. Conf. LINAC94, Tsukuba,
KEK, August 21-26, 1994, Vol.2, pp.786-788.

\bibitem {prl} 
E.G.Bessonov and Kwang-Je Kim, Preprint LBL-37458 UC-414, June 1995;
Phys. Rev. Lett., 1996, vol.76, No 3, p.431.

\bibitem {zhirong} 
Zh. Huang, R.D.Ruth, Phys. Rev. Lett., v.80, No 5, 1998, p. 976.

\bibitem {ICFA01} 
E.G.Bessonov, Proc. of the 23d Int. ICFA Beam Dynamic WS on Laser-Beam
Interactions, Stony Brook, NY, June 11-15, 2001; physics/0111084.

\bibitem {kolom-leb} 
A.A.Kolomensky and A.N.Lebedev, Theory of Cyclic Accelerators. North
Holland Publ., $C^o $, 1966.

\bibitem {bruk} 
H.Bruk, Accelerateurs Circulaires de Particules (Press Universitaires
de France, 1966).

\bibitem {wiedemann} 
H.Wiedemann, Particle Accelerator Physics I \ and II (Springer-Verlag,
New York, 1993).

\bibitem {abb} 
D.F.Alferov, Yu. A.Bashmakov, E.G.Bessonov, Sov. Phys. Tech. Phys.,
v.23 (8), 1978, p. 905.

\bibitem {klotz} 
W.D.Klotz, G.M$\ddot u$lhaupt, Proc. WS on Forth Generation Light
Sources, Febr. 24-27, 1992, Ed. M.Cornacchia and H.Winick, SSRL
92/02, p. 138.

\bibitem {winick} 
H. ~ Winick, Proc. WS on Forth Generation Light Sources,
Febr. 24-27, 1992, Ed. M.Cornacchia and H.Winick, SSRL 92/02, p. 1 -
4; p. 8 - 15.

\bibitem {hara} 
M.Hara, Proc. 10th ICFA Beam Dynamics Panel Workshop " 4th Generation
Light Source ", Grenoble 22 - 25, 1996, p. WG4-42 - WG4-43.

\bibitem {j.chen} 
J.Chen, K.Imasaki, M.Fujita, C.Yamanaka et al., Nucl. Instr. Meth.,
A341 (1994), p.360.

\bibitem {epac96} 
E.G.Bessonov, K.J.Kim, Proc. 5th European Particle Accelerator
Conference, Sitges, Barcelona, 10-14 June 1996, v.2, p. 1196.

\bibitem {esrf96} 
E.G.Bessonov, K.-J.Kim, Sources of Coherent X-rays Based on
Relativistic Ion beams, Proc. 10th ICFA Beam Dynamics Panel Workshop
" 4th Generation Light Source ", Grenoble 22 - 25, 1996, p. W2-108 -
W2-110.

\bibitem {lan} 
Landeker K., Phys. Rev., 1952, V. 86, No 6, P. 852.

\bibitem {lam} 
Lampert M.A. Phys. Rev., 1956, v. 102, No 2, p.299.

\bibitem {aru} 
F.R.Arutyunian, V.A.Tumanian, Phys. Lett. v.4, p.176, 1963.

\bibitem {mil} 
R.H.Milburn, Phys. Rev. Lett. v.10, p.75, 1963.

\bibitem {fiocco} 
Fiocco G., Thompson E., Phys. Rev. Letters, 1963, v.10, p.89.

\bibitem {kulik} 
Kulikov O.F., Telnov Yu.A., Filippov E.I., Yakimenko M.N., JETP, 1964,
'.47, p. 1591; JETP, 1969, v.56, p. 115; Phys. Letters, 1964, v.13, No
4, p.344; JETP letter, 1969, '.9, p. 519.

\bibitem {bem} 
Bemporad C., Milburn R.H., Tanaka N. et al., Phys. Rev. 1965, v. B138,
p.1549.

\bibitem {mikh} 
V.Balakin, A.Mikhailichenko, Preprint BINP 79-85, Novosibirsk, 1979.

\bibitem {besdesy} 
E.G.Bessonov, Proc. 15th Int. Accel. Confer. HEACC92, 1992, Hamburg,
v.1, p. 138; Proc. 6th Int. Workshop on Linear Colliders, March 27-31,
1995, Tsukuba, Japan, v.1. p. 594.

\bibitem {hirose} 
T.Hiroce, Proc. Internat. WS on physics and experiments with
linear colliders, Morioka, Iwate, Japan, September 8-12, 1995,
Singapure, 1996, p.748.

\bibitem {okugi} 
T.Okugi, et al., Jpn. J. Appl. Phys. v.35 (1996), p.367.

\bibitem {frish} 
J.Frish, Proc. of the workshop on new kinds of positron sources for
linear colliders, March 4-7, 1997, SLAC-R-502, p.125.

\bibitem {omori} 
Tsunehiko OMORI, Proc. of the workshop on new kinds of positron
sources for linear colliders, March 4-7, 1997, SLAC-R-502, p.341.

\bibitem {mar} 
Marino A., Matone G., Rocella M., Laser and unconvernt. Opt. J., 1975,
No 55, p.3

\bibitem {glotin} 
F.Glotin, J.M.Ortega, R.Prazeres, et al., Phys. Rev.
Lett., 1997, v.77, No 15, p. 3130.

\bibitem {litv} 
Litvinenko V.N., Shevchenko O.A., Mikhailov S.F.et al., Proc. of 2001
Particle Accelerator Conference, Chicago, IL, USA, June 2001 (in
Seales).

\bibitem {tomi} 
T.Tomimasu, Rev. Sci. Instr., v.60, p.1622 (1989).

\bibitem {csonka} 
P.L.Csonka, R.O.Tatchin, Proc. WS on Forth Generation Light Sources,
Febr. 24-27, 1992, Ed. M.Cornacchia and H.Winick, SSRL 92/02,
p.555-564.

\bibitem {chen} 
Chen J., Imasaki K., Fujita M. et al., Nucl. Instr. Meth., 1994. V. A
341, p.346.


\bibitem {ispirian} 
K.A. ~ Ispirian and A.T. ~ Margarian, Phys. Lett. {\bf
44A}, 377 (1973).

\bibitem {miller} 

L.D. ~ Miller, Opt. Commun. {\bf 30}, 87 (1979);

\bibitem {baldvin} 
G.C. ~ Baldwin and N.J. ~ DiCiacomo,
IEEE Trans. Nucl. Sci.
{ \bf NS-30}, 981 (1983).

\bibitem {basov} 
N.G. ~ Basov, A.N. ~ Oraevsky, B.N. ~ Chichkov, Sov. Phys. JETP, v.62, p.37,
(1985).

\bibitem {glad} 
P.Gladkikh, I.Karnaukhov, S.Kononenko, A.Shcherbakov, NIM, A448, 2000,
p.41; Agafonov A.V., Botman J., Gladkikh P.I.et al., Problems of
Atomic Science and Technology, 2001, No 1, Series: Nuclear Physics
Investigations, p. 126.

\bibitem {vinograd} 
E.G.Bessonov, A.V.Vinogradov, A.G.Touriansky, Pribory i technika
experimenta (in print).

\bibitem {urakawa} 
Urakawa J., Uesaka M., Hasegawa M., et al., Proc. ICFA2001 Beam Dynamic
WS, Stony Brook, NY, USA, 2001 (in a seal).

\bibitem {bes3} 
Yu. Bashmakov, E.Bessonov, Ya. Vazdik, Sov. Phys. Tech. Letters, v.1,
p.239 (1975); Proc. 5th All- Union meeting on charged particle accel.,
"NAUKA", Moscow, 1977, v.1, p. 277.

\bibitem {derb} 
 Ya. S.Derbenev, A.M.Kondratenko,
E.L.Saldin, Nucl. Instr. Meth., v. 165 (1979), p. 201.

\bibitem {clen} 
J.E.Clendenin, SLAC-PUB-8465, 20 July 2000; 9th WS on Advanced
Accelerator Concepts AAC2000, Santa FE, Hilton, June 10-16, 2000,
Edited by P.L.Colestock, S.Kelley, p.563.

\bibitem {zhirong2} 
Zhirong Huang, Proc. 2d ICFA Advanced Accel. WS on the physics of high
brightness Beams (World Scientific, Singapure, 2000), Editors
J.Rosenweig and L.Serafini.

\end {thebibliography}
\end {document}